# Like-minded, like-bodied: How users (18-26) trust online eating and health information

Rachel Xu

Google Jigsaw, rachelxu@google.com

Nhu Le

Gemic, nhu.le@gemic.com

Rebekah Park

Gemic, rebekah.park@gemic.com

Laura Murray

Gemic, laura.murray@gemic.com

This paper investigated the relationship between social media and eating practices amongst 42 internet users aged 18-26. We conducted an ethnography in the US and India to observe how they navigated eating and health information online. We found that participants portrayed themselves online through a vocabulary we have labeled "the good life": performing holistic health by displaying a socially-ideal body. In doing so, participants unconsciously engaged in behaviors of disordered eating while actively eschewing them. They also valued personal testimonies, and readily tested tips from content creators who shared similar beliefs and bodies to them. In doing so, they discarded probabilistic thinking and opened themselves to harm. Our study found that their social media feeds did not unidirectionally influence participants – they also reflected participants' internalized views of health, in an intertwined, non-linear journey. Reducing the online spread of disordered eating practices requires addressing it within young people's social context.

## 1 INTRODUCTION

During the COVID-19 pandemic, reports of eating disorders among youth skyrocketed [1]. Scholars inferred that increased social media usage—which had also increased due to lockdowns, loss of routines, and other hardships—was a key cause of the spike in disordered eating behavior [2]. To better understand the relationship between their social media feeds and their eating and health practices, we set out in 2022 to investigate how youth navigate eating and health information online. In doing so, we built upon previous research into "info sensibility."[1]

We conducted an ethnographic study with 42 internet users aged 18-26 in the US and India to investigate how they navigated online content related to eating and health. We did not seek out participants with formal diagnoses, and instead focused on self-reported experiences with disordered eating habits because we sought to understand eating and health cultures separate from the pathology of diagnoses. Our research questions included: **(RQ1)** How do broader eating-related health cultures shape information journeys? **(RQ2)** What pathways or patterns do young people experience when searching for information about eating and health? **(RQ3)** What information do they trust and why?

---

[1] Info sensibility is a concept that stands in contrast to info literacy. A previous study [24] showed that users value and evaluate content in terms of how much it helps them formulate an opinion or keep up with their peers' interests, rather than valuing and evaluating online content for how factual it is.

Our research yielded several implications:

1. "Healthy vs. Harmful" did not work as a framework for categorizing information, because the same piece of content could trigger one person but help another, or elicit different responses at different times from the same person.

2. We found a coded vocabulary of images and words for disordered eating: the pursuit of "the good life." Participants sought to perform the good life by signaling holistic health and wellness across all areas of life (work, relationships, hobbies, diet, and fitness), while maintaining the illusion that they had not sacrificed anything to achieve it.

3. Participants of color, along with queer and disabled participants, often found communities online where they felt represented, in stark contrast to the spaces they inhabited offline. However, despite online space offering solace and community, many participants confessed feeling they were better off spending less time on their phone, and sought off-ramps from their online good life to the real world.

4. Participants evaluated content through immediate self-experimentation, without conducting additional online searches about the scientific basis behind claims.

5. The scientific basis for claims found online about eating and health rarely mattered for our participants. They assessed trustworthiness through whether a claim felt personally applicable, and whether the messengers were "like-minded" and "like-bodied" as a heuristic.

We have structured the paper in four parts: a literature review, methodology, key results, and conclusion.

## 2 LITERATURE REVIEW

For the past 40 years, there has been broad agreement that etiological models of eating disorders "should be multidimensional" [3]. Clinicians, theorists, and researchers agree that there is no single cause or set of causes to which disordered eating behavior can be traced [4, 5]. Sociocultural influences such as media exposure, pressure for thinness, and thin-ideal internalization, for example, can exacerbate – or be exacerbated by – personal characteristics (negative emotionality, perfectionism, or negative urgency) [6].

Efforts to factor in the role of culture in biological models of eating disorders often reduce culture to a "risk factor." Culture is most often used, in fact, as a reductive "proxy for racial or ethnic groups" [7]. According to medical anthropologist Rebecca Lester, research that approaches culture as a static variable is not entirely without value.[2] But in focusing on cultural membership as a risk factor and on individuals as passive recipients of media images, these approaches cannot account for how people use practices around eating and health–including disordered eating habits– to actively make meaning in their lives. To get at these deeper questions linking behavior to belief, identity, and aspiration, we need to explore "how the practices of body discipline and affective modulation shape a person's sense of being in the world" [8].

From sub-cultures of carnivory to warring factions of vegan YouTubers, there are a growing number of cultural identities available to the mainstream [9, 10]. A growing number of online cultures, subcultures, and countercultures exist to help people navigate the terrain related to health and eating [11, 12]. Currently, however, we lack a robust literature about the proliferation of health-related identities among those 18-26 years of age to match the comparatively well-established body of research on young people's "networked" lives [13, 14, 15, 16]. We sought to understand how new identities–and the sense of community affiliation they make possible–were mediated by online information. We conducted this research during the COVID-19 pandemic, which saw a rise in reports of disordered eating behaviors among young people [1], that were linked to lockdowns, uncertainties, and increased social media usage. Other media theorists have noted that with the increased digitization of social life with the pandemic, teens and young adults have worked out new ways to care for each other through online spaces [17].

---

2 Interview conducted on June 30, 2022 with Rebecca Lester

Academic debates on internet usage and eating disorders have frequently focused on ethical questions regarding "pro-anorexia" ("pro-ana") websites and forums which portray extreme fasting as a lifestyle choice. Not all dietary misinformation, however, can be as easily identified. "Thinspo," or content that seeks to motivate the user to work towards thinness, has been shown "to have a negative impact on viewers with regard to body image…even five minutes of exposure to thin and beautiful images" can negatively affect body image [18]. Basing their research on a survey of pro-ana internet searches and websites, Lewis and Arbuthnot argue that "search terms with references to thinspiration and thinspo are associated with the most harmful Website content" [19]. Some content may then be deemed harmful by its ability to send users into a "toxic spiral" where content becomes intensified and more explicitly harmful [20].

At the same time, however, there is still a limited understanding of how and when tech companies can successfully mediate a user's informational health ecosystem. Trigger warnings and censorship, for example, have not historically worked [21, 22]. Moreover, critics have argued that dieting apps that use an educational model fail to understand that poor health decision-making does not come from lack of knowledge, but from social stratifications and differing definitions of health [23]. Such critiques highlight the gap between conventional paradigms of information literacy and what users 18-26 actually do online [24].

## 3 METHODOLOGY

We conducted a qualitative research study from June to October 2022 with 42 participants engaged in eating and health online content, between 18-26 years old, evenly split between those from the US and India. We intentionally did not recruit participants with "official" diagnoses of an eating disorder from a medical professional. Instead, we relied upon self-reported experiences with disordered eating and recovery. We sought to understand eating and health cultures separate from the pathology of disordered eating.

**Research ethics**

We solicited participants' informed written consent during the recruitment process and confirmed their consent before in-person ethnography commenced. A licensed psychologist specializing in eating disorders reviewed all prospective participants' answers to our screener, to confirm that we excluded participants whose responses indicated risk for distress. Together with this expert, we designed a de-escalation procedure in the event that a participant exhibited distress or disclosed sensitive medical details. We emphasized the participants' agency to end the interview, pivot away from certain topics, or pause the video recording at any time. We maintained strict data privacy protocols for all participants. All data collected, including field notes, used only pseudonyms assigned to each participant. All participants were instructed to withhold personally identifying information (PII). Raw interview and diary data was scrubbed of PII during transcription. To mitigate risk of participant reidentification, we have omitted personally identifying details or phrases from quotes in this paper. Pseudonyms used below are not the same pseudonyms used in data collection.

**Recruitment**

In the US, we selected participants through the platform respondent.io. We chose participants from the broader New York area: 11 urban, 5 suburban, 5 rural. Our participants included: 2 Asian, 4 Black, 4 Latino, 11 White; as well as: 10 female, 5 male, 4 trans, 2 nonbinary. In India, we used a local recruiter, and included participants in Delhi and nearby peripheral areas: 15 urban, 3 semi-urban, and 3 semi-rural. We included 10 men and 11 women. We also adapted our approach and language for recruitment and research in India to account for cultural norms around food and eating.

**Research Methods**

Our study had four phases of data collection.

    *Phase 1 Literature review and diary study:* We conducted a literature review and expert interviews[3] to understand how disordered eating has evolved in the age of technology, and how the two geographies differ culturally and historically when it comes to eating and health. We also solicited from selected participants a collage of images that reflected their ideas of what is "healthy" and "unhealthy." These collages were used as a starting point for conversations about their information sources and their personal assessments of their own health.

    *Phase 2 In-person ethnography:* We met with participants in their homes or in their neighborhoods, and we followed them on an eating-or health-related activity of their choosing, such as grocery shopping trips or jogging in the park. Over the course of 90 minutes, researchers probed into participants' 1) life histories; 2) information ecosystems; and 3) experience with disordered eating (if applicable).

    *Phase 3 Data synthesis and remote follow-up interviews:* Following the in-person ethnography, the research team identified an initial set of participant needs. We workshopped these with other researchers, product managers, and policy specialists from Google working on issues of health and user safety, to discuss potential interventions. We shared our workshop discussions, including intervention concepts, in follow-up interviews with participants to solicit more qualitative data and to validate our interpretations of their experiences (i.e. member checking).

**Data Analysis**

Preliminary analysis began immediately after the in-person ethnography based on field notes. The research team discussed the interview and diary study data after watching the recordings and reading participants' textual responses, and identified insights related to our RQs. This allowed us to tailor the follow-up remote interviews to each participant. During both rounds of interviews, we coded collected data iteratively, yielding 150+ discrete insights. We used the coding to assess the saturation of insight across the dataset as a whole and among specific subsets of respondents. After data collection and coding were complete, we iteratively clustered insights into themes, and created rough thematic maps. We did this across several collaborative and independent analysis sessions.

**Limitations**

Our study's sample was not statistically representative of the broader population. We limited the study to two geographic locations accessible to urban areas. However, we sought to diversify the sample because we were conscious of historical biases of research into disordered eating (focusing on participants who were predominantly white, wealthy, and female). We deliberately included participants who were male, queer, and non-white. We also chose to diversify our sample across geographic density, income, and eating and health behaviors–however, there may have been other characteristics (e.g. disability[4]) that may have been relevant to eating and health cultures.

---

3 Expert interviews with the following academics: Hanna Garth, Rebecca Lester, Krishnendu Ray, Lesley Jo Weaver.

## 4 RESULTS & DISCUSSION

**"The Good Life"**

In our diary study, we asked participants to create two collages of images found on the internet that represent "healthy" and "unhealthy." However, we quickly discovered that the same content could be helpful or harmful depending on the person, or where the participant was in their journey with eating and health. All of the participants internalized ideas about eating and health at home, from their families, and when their attitudes and behaviors changed over time, they felt that their social media feeds reflected, not caused, their internal responses to content around bodies, dieting, and exercise. For example, Jaime (23, US) explained that prior to adopting a "recovery mindset," content intended as helpful by others was often experienced by Jaime as harmful:

> "[For me] this recovery stuff is more triggering and instructive... When I was 12, it helped me learn how people were engaging in [anorexic] behaviors."

Today, Jaime can view and critique pro-ana content and not be triggered by it. Jaime's experience was shared by many participants: content was not categorically wholesale harmful or helpful. Content unremarkable or helpful to one person can be triggering for someone else. Thus, rather than categorizing content, we focused instead on understanding eating and health cultures online and their effect on our participants' relationship to eating and health.

The broader picture that formed in our fieldwork in both the US and India was a common visual and textual vocabulary shared among our participants that we refer to as "the good life." By vocabulary, we are referring to not only words, but also images, videos, and discourses—all of which coalesce around broad definitions of wellness, including content that arguably encourages disordered eating tendencies or recycles the same content that was once labeled as pro-ana. While no participants used the phrase "the good life" explicitly, we observed that in both cultures, all of our participants held the notion that they could achieve "goodness" through healthy eating and exercise. The ideas of "goodness" and "badness" underscored their entire language, beliefs, and cultures around eating and health, and this pursuit of being seen as good in one's cultural context proliferated through eating, exercise, and lifestyle content. Although in reality being a good person could manifest in many ways (including kindness, altruism etc.), for our participants, the body was a key marker and proxy of how good one is living life. As Frida (24, US) succinctly put it:

> "In the back of my mind, I thought to be good, you have to be skinny."

While examinations of online discussions of disordered eating often focus on those which are explicit and pathologized [25, 26], we found our participants using words such as "energy," "mental health," and "feeling good" to describe their eating and health journeys online, even when recounting restriction and binge cycles. Living the good life meant looking "effortlessly good" and "leading a rich life beyond work." Participants pursued diets and exercise routines for the sake of "self-improvement", "mental health", or "balance", which were all perceived to be socially acceptable goals, in contrast to wanting to explicitly lose weight and be thin. In the US, the discourse revolved around mental health as taking care of physical health was often done in the service of mental health. In India, the discourse revolved around being "balanced". However, participants' social media feeds and practices reinforced the idea that only certain bodies fit the aesthetics of the good life: thin bodies for women, and muscular bodies for men. Many of our participants engaged in harmful disordered eating cycles while pursuing and performing the good life.

---

[4] We had one disabled participant in our study and their nonbinary experience could have been explored more widely with other participants with a range of disabilities.

It was not enough to simply live the good life–participants also felt compelled to broadcast online that they had attained it, and done so with minimal effort. For example, Allie (24, USA), a self-described overachiever, medical student, and community builder, experimented with various fad diets. However, she did not call it dieting, instead saying she wanted to be "healthy across the board." Allie's Instagram page was carefully curated to project a lifestyle of ease and health, for example, a "candid pose" in an outfit that took many attempts to best highlight the physical results of her diets. In the US, social media was most commonly used by our participants to showcase success in all domains. In India, social media provided our participants with access to global norms and practices (e.g., steroid use), filtered through localized influencers who were also posting similar types of content within this shared vocabulary around the good life.

Seeking to change their bodies is not an irrational exercise; our participants were acutely aware that having socially-ideal bodies came with access to real social benefits. For example, Bailey (24, USA) was moved up to the front of their dance class after losing weight. Abdullah (19, India) said, "Everyone is attracted to the steroid-type body," where muscles were visually associated with success in career and economic status. In both the US and India, the body is perceived as grounds for social judgment, affecting what one is granted or denied.

Our findings around the good life connect with the work of anthropologist Rebecca Lester, who theorized disordered eating behaviors are not about pathological fear of food and eating, but rather a yearning to be perceived as a good and valuable person in the world, expressed through the body as the fundamental unit of identity–which Lester described as the "deep why".[5]

It is worth noting that, despite being steeped in the good life vocabulary online, an emerging element of living a life that was truly fulfilling was having a moderate approach to social media and screentime. Living and performing the good life on their phone ultimately didn't feel like a truly fulfilling life, and all of our participants sought off ramps from online worlds into "real life." Nihal (21, India) shared:

> "During the lockdown, I felt no energy and spent so much time on my phone late at night. I had a 'self-realization' that I needed to get out there and change my life."

Yet, we observed that the way participants spoke about the binge-purge tendencies around food was similar to how they tried to manage their social media usage. They did not moderate or taper their screen time. Instead, just as they would cut out certain foods for a couple of weeks to lose weight only to then give up the restrictive diet, participants would delete apps only to reinstall them. While they often yearned for alternatives to the online good life, few could resist eventually returning to it. This seeming inescapability of the good life is detrimental to our participants because the good life did not challenge messages connected to disordered eating. In fact, the good life vocabulary obfuscates harmful discussions around eating disorders, disguising this content as healthy aspirations to both platforms and users.

**"N=1 Thinking"**

Our participants overwhelmingly preferred eating and health information from "real" experiences of individuals online over information from institutions or perceived experts. This mirrors our previous findings that young people (18-26) value personal testimony over institutional authority [24]. Many of our participants think about eating and health through the lens of identity – and particularly those who identify with marginalized or emergent identities were wary of how science has historically harmed and pathologized marginalized people. Within this context, we observed a practice that Krishnendu Ray describes as "N=1 thinking," where people regard personal testimony as valid evidence. Surya (20, India) shared:

> "If someone is suggesting something to you and it has worked for them, it gives a little bit more authenticity to it... I don't want experts and their advanced workouts, you don't know which one is authentic or if it will work for you."

---

5 Interview conducted on June 30, 2022 with Rebecca Lester.

Participants' faith in the validity of N=1 thinking was bolstered by the fact that participants were often not watching just one testimonial, but a string of them at a time. Enough repetition of the same kind of testimony increased their perception that the claims being made were true [27, 28]. For example, Jamuna (23, India) trusted the bhangra dancer diet and exercise tips she heard precisely because they were repeated across many different bhangra influencers.

In shaping their eating and health practices, we found that participants were comfortable favoring personal testimonies over scientific studies because they were not looking to become experts on the topic of health, but rather to become experts on their own bodies. Even if a diet or workout suggested by a creator did not work for them, participants did not discount it–they simply assumed it was not right for their bodies, not that the information itself was flawed. Aadita (23, India) described her experience following diets on YouTube and Instagram:

> "I didn't lose 10 kgs in 10 days, I lost 2 kgs, so it just didn't work for me, but it could work for somebody else."

The lack of probabilistic thinking around health content in favor of N=1 thinking is problematic because it ensures that online testimony (which could contain harmful advice) cannot be proven false, because it is about experience, not facts. When watching a video about someone else's new diet, Laney (23, USA) shared that she does not worry if that person's explanation does not sound plausible or stand up to scrutiny. She explained:

> "I might not believe the science behind it–your explanation– but I believe your experience is legit. I accept that some things just don't have an explanation."

From our participants' perspective, if institutions are untrustworthy, and anyone's experience could theoretically be plausible, the only way to know if something works for them is to try it on themselves–which all of our participants did regularly, without vetting information first. Use of N=1 thinking especially as it relates to eating and health means that young people are at risk of experimenting on their own bodies, which may have unintended and lasting consequences on their health and well-being.

**Like-minded, Like-bodied**

Our participants sought out information on eating and health delivered by social media creators who were similar to them; or more specifically, both "like-minded" and "like-bodied." Like-mindedness is when the user and health-content creator share similar reference points, language, backgrounds, humor, or politics. Like-bodiedness is when the user and health content creator share similar body types, and look to be of the same age and the same race/ethnicity. Like-bodiedness connects people who are assumed to have similar genetics, health journeys, and responses to diet and fitness trends in the past. Participants sought out such creators both consciously, by active searching, and unconsciously, by staying on suggested content in their social media feeds.

Pursuit of the like-minded and like-bodiedness was one way participants could find information that was relevant or personalized, and underscored all of their information-seeking behaviors for health and eating content. Participants were particularly compelled by good life vocabularies produced by like-minded like-bodied individuals, because it was theoretically attainable by someone who looked and behaved like them. For example, Zuri (26, US) enjoyed Black cottage core, which represented a modern version of "the little house on the prairie," because it was "therapeutic" to see aspirational images of Black influencers. When exercising N=1 thinking, participants assigned particular weight to testimonies from the like-minded, like-bodied. Participants believed that shared similarities raised the chances that a fitness regime would work for them. For example, Janet (22, USA) said that because she had a bigger body, she liked following a fitness influencer who had her body type.

Participants sought out like-minded, like-bodied messengers to build supportive communities online, often motivated by what they lack in the real-world. Many of our participants, particularly those who felt marginalized in the real world (participants who were people of color, transgender, and/or disabled) found solace in like-minded, like-bodied people and communities online, where they felt included. For example, Janet (22, US) never felt valued in her hometown, which she describes as predominantly "white and skinny," instead finding solace in online creators who had bigger bodies. In India in particular, YouTube was becoming a home for localized voices on health and eating – where "local" became a useful proxy for like-minded like-bodiedness.

But as powerful as it was to see themselves represented online, ultimately our participants all shared an instinct that online communities could not completely substitute for real world engagement, experiences, and connections. Participants were often experiencing the tension between a more appealing online world, where connecting with others felt easier, and the real world, where people who were like them were not all neatly gathered in one place and where social anxieties abounded.

**CONCLUSION**

Our participants have developed a vocabulary for discussing and disguising disordered eating through the pursuit of the good life. Their feeds were filled with like-minded and like-bodied content, and they accepted personal testimonies on social media platforms as trustworthy and valuable (N=1 thinking). Participants ultimately evaluated content through self-experimentation. Our findings suggest that to truly address disordered eating on social media, we must understand that personal journeys and information journeys exist in parallel in a symbiotic relationship. One of our key takeaways is that social media is a reflection of cultural views on disordered eating and the good life, and not solely a driver as previous studies have often argued. Social media can certainly intensify and spread viewpoints and practices of disordered eating, but addressing it in isolation from the rest of a participant's cultural context does not address the underlying "deep why." Our goal was to show how the desire to be culturally recognized as a good person played out among our participants, and how this materialized through the pursuit of the good life both online and offline.

**ACKNOWLEDGMENTS**

We extend our deepest gratitude to Amelia Hassoun, Todd Carmody, Devika Kumar, Beth Goldberg, Suzannah Isgett, Yasmin Green, and Shira Almeleh for their many efforts in support of this project. We would also like to thank our experts Hannah Garth, Rebecca Lester, Krishnendu Ray, and Lesley Jo Weaver for sharing their knowledge. Finally, we thank our participants for sharing their experiences with us.